# Please Don't Go: Understanding Turnover of Software Engineers from Different Perspectives

MICHELLE LARISSA LUCIANO CARVALHO*, PAULO DA SILVA CRUZ*, and EDUARDO SANTANA DE ALMEIDA, Federal University of Bahia (UFBA), Institute of Computing (IC), Brazil
PAULO ANSELMO DA MOTA SILVEIRA NETO, Federal Rural University of Pernambuco (UFRPE), Computer Institute, Brazil
RAFAEL PRIKLADNICKI, School of Technology - PUCRS, Brazil

Turnover consists of moving into and out of professional employees in the company in a given period. Such a phenomenon significantly impacts the software industry since it generates knowledge loss, delays in the schedule, and increased costs in the final project. Despite the efforts made by researchers and professionals to minimize the turnover, more studies are needed to understand the motivation that drives Software Engineers to leave their jobs and the main strategies CEOs adopt to retain these professionals in software development companies. In this paper, we contribute a mixed-methods study involving semi-structured interviews with Software Engineers and CEOs to obtain a wider opinion of these professionals about turnover and a subsequent validation survey with additional Software Engineers to check and refine the insights from interviews. In studying such aspects, we identified 19 different reasons for Software Engineers' turnover and 18 more efficient strategies used in the software development industry aiming to reduce it. Our findings provide several implications for industry and academia, which can drive future research.

CCS Concepts: • **Social and professional topics** → **Computing occupations**; **Employment issues**.

Additional Key Words and Phrases: Turnover, Software Engineers, CEOs, interviews, survey

## 1 INTRODUCTION

Due to the evolution of technology and the rise of the internet, web-based businesses and the expansion of the information economy have challenged organizations to seek, recruit, and retain Information Technology (IT) professionals such as software developers [3]. Such professionals are in a highly interconnected world [18], where their jobs can be performed many times from anywhere, and they can easily move from one company to another [56]. However, based on this flexibility

---

*Both authors contributed equally to this research.

Authors' addresses: Michelle Larissa Luciano Carvalho, michellellc@ufba.br; Paulo da Silva Cruz, cruz.paulo@ufba.br; Eduardo Santana de Almeida, esa@rise.com.br, Federal University of Bahia (UFBA), Institute of Computing (IC), Salvador, Bahia, Brazil; Paulo Anselmo da Mota Silveira Neto, Federal Rural University of Pernambuco (UFRPE), Computer Institute, Recife, Brazil, paulo.motant@ufrpe.br; Rafael Prikladnicki, School of Technology - PUCRS, Porto Alegre, Brazil, rafaelp@pucrs.br.

and the short tenure of these professionals in companies, it generates a phenomenon known as turnover [40]. Turnover consists of movement coming into and out of professional employees in the company in a given period [58]. This phenomenon is voluntary when the employee himself decides to leave the company. In contrast, it is involuntary when the leadership decides that the employee anymore does not add value to the company [1].

We have noticed a negative impact on the organization in terms of turnover. Such a phenomenon caused in software development companies generates knowledge loss in software projects, delays in the project schedule, and increased cost in the final project [10, 29, 45, 65]. In general, the high turnover rates usually trigger disruption of operations, resulting in the company's declining performance and productivity [15, 22]. Aiming to meet the increasing demand of the software development industry, software production has come to have a significant share of the intellectual capital, which is centralized mainly in the Software Engineer role [26, 39, 59]. However, software companies have been facing difficulties retaining such professionals for a long period [6]. Indeed, organizations are challenged to create effective recruitment and selection strategies to retain Software Engineers on the job vacancies [3].

Although there are more than 26.9 million Software Engineers worldwide, it is insufficient to meet the software industry due to the continuous expansion of technology and the growing global shortage of professionals in this field [47]. Faced with this scenario, organizations are concerned with fostering actions to provide development and retention of these professionals. Success depends on retaining the company's talents by avoiding losing them to competing organizations [5]. Voluntary turnover is high in technology companies (about 69.9%), which has worried companies and CEOs [27]. Once Software Engineers are considered a key part of the software industry, it is necessary to conduct more studies to understand the contributing causes of turnover by considering personal and organizational aspects. In addition, it is important to identify companies' strategies to retain such talents [39].

Existing studies in the literature report problems concerned with staff turnover and recommendations on how to retain IT professionals [9, 15, 19, 35, 46]. Additionally, Nunes *et al.* [46] and Frufrek *et al.* [19] identified strategies to reduce turnover. However, these studies analyze only aspects from the organizational point of view, such as human resources strategies in hiring, job satisfaction, the relationship among colleagues, and organizational commitment [56]. Therefore, we identified a few details regarding the aspects that influence turnover. Moreover, the studies are not directed at Software Engineers. This paper contributes to turnover with two closely related goals in this context. Our first goal is to understand the motivation that drives Software Engineers to leave their jobs and how some personal aspects can influence this decision. Our second goal is to understand the other side of the problem: the main strategies adopted by CEOs to retain these professionals in software development companies.

We conducted semi-structured interviews with 11 Software Engineers and 8 CEOs from different companies. We then surveyed 326 Software Engineers across twenty-five countries and four continents to contextualize and augment the interview findings. Our focus in conducting these interviews and surveys was on the two goals mentioned above.

The following two research questions guided the analysis of the interviews and survey responses:

- **RQ1. What are the main reasons for Software Engineer's turnover?** Several aspects can cause software professionals to leave a company. We seek to identify the main reasons for the turnover of Software Engineers in the software development industry.
- **RQ2. What are CEOs' most effective strategies to mitigate Software Engineer turnover?** We seek to identify the opinion of Software Engineers about the most efficient retention strategies software development companies use.

This paper constitutes the first broad empirical study that reports an understanding of the Software Engineers' turnover behaviors in the industry. It makes the following contributions:

- A mixed qualitative and quantitative study that investigates the Software Engineers' turnover phenomenon in the industry.
- A set of twelve observations regarding Software Engineers' and CEOs' perceptions of turnover in the software industry.
- A collection of all our research materials on a project website for replication and reproducible purposes, including our interview data (prompts, transcriptions, and codebook) and the survey instrument.

## 2 TURNOVER: AN OVERVIEW

This section presents the definitions of turnover, its theories, and models that have served as the basis for current studies. In addition, it presents the causes of turnover and strategies that companies use to reduce it.

### 2.1 Definition

Turnover is one of the main problems that currently concerns professionals in the human resources of organizations [12]. The high level of turnover directly affects productivity, profitability, and organizational health since it is considered the suspension of an organization's membership by an individual who received monetary compensation [22, 43]. Mobley *et al.* [43] point out that the organization must identify the main determinants of such phenomenon and take the necessary measures to prevent it from becoming costly. In their opinion, exit interviews with people leaving the job can be considered a tool to gather information and diagnose the causes of turnover.

The loss of an employee is a critical aspect of the organization, and the difficulty in finding the right talent to meet the demands is not always quick and easy [25]. In general, it takes time for the new employee to achieve high performance, resulting in a loss of competitiveness in the market [1, 13]. However, voluntary turnover causes negative consequences for the organization, such as project schedule delays and productivity loss. In contrast, involuntary turnover can be beneficial because an employee with low productivity will be replaced by another with better productivity [1].

Theories and models have been used to explain turnover by analyzing organizational behavior. Although some of these theories are considered old, nowadays, they are used as a basis for several studies [43]. The oldest theory is named *The Organizational Balance Theory*, which points out that the reason for turnover is employees' dissatisfaction when they perceive that their contributions are more significant than the incentive offered by the organization [23]. Another existing theory is named *The Expectations Met Theory*, which is based on the organizational balance theory of turnover. According to Porter *et al.* [49], meeting expectations determines turnover decisions. In contrast, *The Linkage Model* suggests a series of possible mediation steps between dissatisfaction and actual quitting of a job [42].

Other existing models address the motivations of the employee to leave the organization. *The deployment model*, for example, addresses what drives the employee not to leave the organization [34], whereas *Rooted Turnover Theory* states that the employees are rooted in a kind of web, which prevents them from leaving their job [41]. *The Contagion Model of Turnover* theorizes that co-workers influence the individual's decision to leave or stay in a job, *i.e.*, one co-worker search for job alternatives or actual quitting can spread through a process of social contagion in which affects the behavior of another quitting employee [16].

## 2.2 Causes

Organizations should act on the causes of turnover instead of acting on their effects [12]. In this sense, it is essential to diagnose the causes of turnover and its determinants by considering the adoption of new policies for human resources, which is of paramount importance. Several internal and external phenomena cause turnover; however, we can highlight *the lack of organizational commitment* as the leading cause. When the employee is more committed to the organization, they are more productive and satisfied staying at the job [63].

Ghosh *et al.* [20] consider *the lack of professional recognition* a cause for turnover. The *lack of salary policy* and *career plans* also strongly cause turnover. Such incentives, when applied, can maintain intellectual capital within the company by avoiding turnover. Similarly, socialization is the primary practice for integrating employees when a negative relationship with co-workers exists. The *bad relationship factor* can lead to stress at work, influencing the employee's decision to leave the company [49].

According to Joseph *et al.* [30], an *inadequate work environment* strongly correlates with dissatisfaction and stress at work. When an employee feels discomfort performing his duties, it is replaced by the intention and the desire to leave. In addition, the *trade-off between work and family* is an external factor since it is also linked to *stress* and causes the intention to leave [10]. A *bad boss* with a negative organizational culture, who is very authoritarian or liberal, and who deals with non-appreciation without continuous feedback is also a factor that has a strong influence and leads the employee to leave the organization. Therefore, the *inadequate managerial style* is an organizational factor that causes turnover [36].

Additionally, the repetitive work and exposition to an exhaustive work overload negatively influence employees' satisfaction, thus leading such a professional to practice turnover [10]. The lack of communication in the workplace is also an antecedent of exhaustion. Therefore, turnover is also caused when is not established clear, objective, and spontaneous communication with team members [44]. In general, the factors for turnover are multi-causal and related to the organization as well as external and personal factors. Once turnover occurs, the effects and consequences in the organization range from productivity loss to project delivery delays. In this sense, it is of paramount importance to know the impact of turnover on organizations.

## 2.3 Effects

The effects of turnover in a software development project can trigger delays in the delivery schedule and financial losses. This occurs because the entire project depends on intellectual capital [10]. In large software development projects, tacit knowledge is threatened by developer turnover. In this sense, training is needed for the newly hired employee to understand the business model and other aspects of the organization [50].

In addition, the unscheduled departure of an employee disrupts organizational processes and reduces productivity. In this context, the production will only be normalized when a new employee can produce fully [10, 20]. Additionally, aiming to recruit new employees, the company has a direct cost related to the disclosure of vacancies, selection, and finally, training [3]. The whole process of hiring new employees leads to an overload of other employees who end up assuming the obligations of the employee who left. Therefore, this increases the project's final cost due to new hires and training [10, 29]. Thus, organizations must create strategies to prevent or minimize it from occurring to avoid turnover.

## 2.4 Strategies to Minimize Turnover

As previously discussed, employee turnover can negatively affect the company when poorly managed. Project managers must define strategies to control such a phenomenon and, thus, minimize the risks of project failures. In this sense, effectively identifying risks can proactively reduce and prevent turnover, benefiting organizations in tangible and intangible ways [33, 52]. Strategies from the financial and hiring perspectives can contribute to the organization's employees' stability [60].

Regarding Stovel *et al.* [60], it is necessary to pay attention to transforming work developed by human capital into knowledge and skill. In their study, the authors found that knowledge assets are not effectively managed and that voluntary turnover is a foe of knowledge management. Therefore, managers need to be concerned about the implications of losing valuable employees to the competition and be motivated to minimize the effects of this turnover. This is possible only using *turnover contingency planning* and *knowledge management strategies*. With knowledge management strategies, companies can leverage their employees' experience to create future solid business plans.

Nunes *et al.* [46] reported good practices to be used in companies to reduce turnover such as *concern with the workplace*, *honesty and organizational transparency*, *compensation*, *offering good benefits*, *offering professional growth*, *providing challenges for employees*, recognizing and training, and *managerial style and autonomy*. The findings presented by Frufrek *et al.* [19] indicated that the *lack of professional recognition*, *lack of salary policy*, *lack of organization involvement*, and *personal expectations* influence the professionals' decision to leave the company and can contribute decisively to staff turnover rates. In addition, a performance-based benefits policy creates a sense of motivation as long as it is distributed fairly. In the same way, *a salary aligned with the labor market* and *a career plan aligned with the possibility of professional growth* provide commitment and better satisfaction with the company, contributing to the reduction of turnover [3, 20].

## 3 RELATED WORK

Academics and professionals have studied the turnover phenomenon for decades since it is a highly relevant and vital issue for companies in all sectors [2]. The existing studies focus on predictors of turnover at the individual level, employee demographics, job satisfaction, and organizational commitment [11].

Hom *et al.* [24] presented a review of publications on employee turnover covering the 100-year existence of the Journal of Applied Psychology. Such a study discussed significant theoretical and methodological contributions to the literature on this phenomenon. However, as the turnover study is dynamic and constantly changing [24], it challenges us to seek new understandings of turnover from different personal and organizational perspectives.

Employee turnover significantly impacts the software industry due to dependence on intellectual capital once professionals, after hiring, take a while to become productive [8, 59]. Thus, understanding the professional turnover in Software Engineering is very important. Farooq *et al.* [15] conducted a literature review study to identify turnover factors. For an empirical assessment of these factors, the authors surveyed four hundred and forty IT and software professionals and analyzed the findings using linear regression. However, the research was limited to just one country, Pakistan. The results showed that *recruitment and selection*, *management support*, *performance*, and *career management*, *salary and remuneration*, *employee commitment*, *job security*, and *recognition that measure job satisfaction* significantly impact turnover intention. In addition, personal demographics such as gender, age, level of experience, and work experience were also evaluated. This analysis was based on job satisfaction as a mediator with turnover intention, which resulted in non-significance for age and gender concerning the impact on IT professionals' turnover intentions.

Buhari *et al.* [9] identified turnover factors based on the literature review and surveyed ninety-six Software Engineers to validate them. The authors conducted a study in Siri Lanka and examined job satisfaction as the primary mediator. The study results showed that professional commitment positively influences job satisfaction and also decreases one's intention turnover. In this sense, IT professionals seek satisfaction related to co-workers, supervision, and work design. On the other hand, perceived organizational support did not significantly affect the intention to leave the organization. In addition, there was no difference by gender or career stage in the effect of job satisfaction on turnover intention.

Sharma *et al.* [56] presented a theoretical model targeting the software engineering domain to help organizations understand which factors play a role in achieving success in the onboarding process for software professionals and how this might impact job satisfaction workplace relationships, and reduction of the level of turnover. The theoretical model analyses new hires using job satisfaction and job quality as an intermediary to measure short-term and long-term intentions of leaving or remaining in the organization. To evaluate it, the authors surveyed one hundred-two software professionals. The study results showed that support for new hires is essential in successful integration but that training is less important. In addition, job satisfaction intermediates the relationship between integration success and turnover intention.

Unlike the previous studies, Nunes *et al.* [46] conducted a study in Brazil based on a structured survey with nineteen questions focusing on managers of software development companies to identify the most effective strategies for employee retention. Based on a literature review and the study with forty-six managers, the authors identified factors regarding turnover and strategies to reduce such a phenomenon. Frufrek *et al.* [19] also applied a survey to determine the causes and effects of employee turnover in the software industry in Brazil. Based on this survey, it also was possible to investigate which strategies are applied to minimize such a phenomenon and its effects. They analyzed the findings from sixty-seven software professionals using descriptive and inferential statistics.

Although the studies mentioned earlier focus on software professionals as the target audience [9, 15, 19, 46, 56], none of them focus on evaluating the turnover reasons and most effective strategies that can be used to mitigate turnover from the perspective of Software Engineers and CEOs of the software development industry. In general, software development companies should be concerned with factors that can motivate employees to stay and use efficient strategies to minimize the effects of turnover. The existing literature suggests some of these aspects that can be used to deal with this phenomenon, such as *work environment, adoption of salary policy and career plan, adoption of a performance-based benefits and promotions policy, documentation of project activities to assist new employees, conducting periodic training in new technologies, performing activities in pairs, role alternation for the employee to perform more than one function, using interviews to assess risk before hiring*, and *conducting training at the beginning of employment* [19]. However, just the studies presented by Nunes *et al.* [46] and Frufrek *et al.* [19] surveyed software development companies to identify which strategies are applied to minimize employee turnover and its effects.

Although personal demographic data has been considered in the presented studies, a deep analysis of these aspects is not evident, and most research reveals no clear correlation among them. Furthermore, these studies are limited to their country of origin, making the findings harder to generalize. Indeed, it is essential to consider that IT professionals are not part of a homogeneous group since, depending on the type of work, they differ in their personality, work-related attitudes, and turnover intentions [21]. Our study aims to fill these gaps by investigating how Software Engineers perceive the factors causing turnover and which retention strategies are most effective in mitigating such a phenomenon. Besides performing a literature review, we conducted semi-structured interviews with Software Engineers and CEOs of software development companies to collect and better understand reasons for Software Engineers' turnover and strategies used by CEOS to mitigate it. Moreover, we surveyed software engineers from different countries to augment and validate the interview findings. We also assessed the correlation between personal demographic factors such as *age*, *gender*, *level of experience*, *generation*, and *personality types* and their impact on the decision of Software Engineers to leave the company.

Fig. 1. Research Design approach.

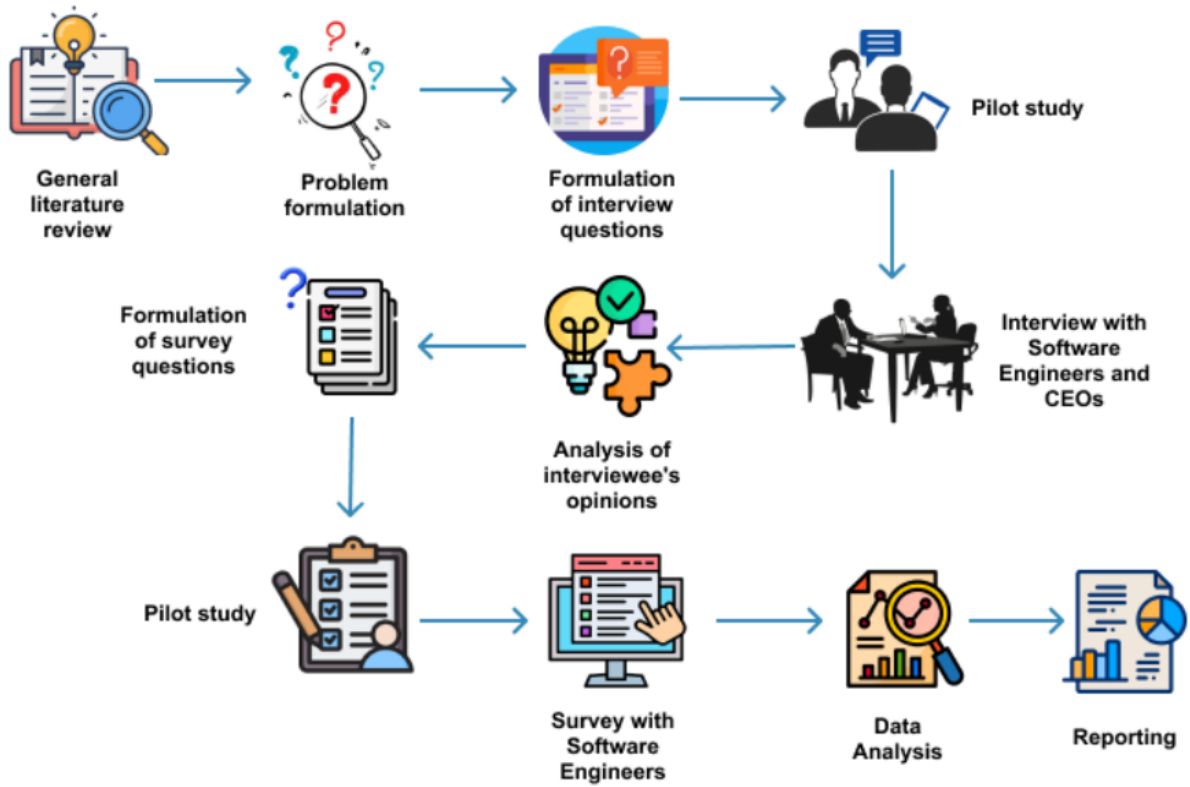

## 4 RESEARCH DESIGN

This section describes the research design used in our study. Firstly, we *(i)* reviewed the literature to understand better the causes and consequences arising from turnover, besides the strategies used by companies to minimize it. Next, we *(ii)* conducted in-depth semi-structured interviews with Software Engineers and CEOs to obtain a wider opinion of these professionals about turnover. Finally, we *(iii)* performed a validation survey with additional Software Engineers to check and refine the insights from interviews. Figure 1 shows the overall research design used in our study. All study materials from interviews and surveys can be found in the supplementary materials for the paper[1].

### 4.1 Interviews

**Protocol.** The interviews were conducted firstly with the Software Engineers and next with the CEOs. We interviewed 11 Software Engineers and 8 CEOs from industry. The average length of the interviews was 26 minutes, the median of 24.5, the shortest 14, and the longest 42 for Software Engineers. In contrast, for CEOs, the average length of the interviews was 26.9 minutes, the median was 30, the shortest was 15, and the longest was 36 minutes. The interviews were conducted remotely via the Google Meet platform[2]. It allowed for recording the interviews and facilitating later transcription.

---
[1]https://github.com/software-engineering-turnover/empirical-study

We previously conducted pilot interviews to validate the research instrument, where the Software Engineer was identified as P1 and the CEO as C1. Moreover, we provided an adequate explanation of the research so that the interviewee could understand the study's objectives and had a more active participation [55]. In addition, we conducted interviews between June 2022 and September 2022 with Software Engineers and from October 2022 to December 2022 with CEOs. The interviewer asked the questions during the interviews, and then the interviewee answered without interruptions. The interview prompt comprises ten questions for the Software Engineers and CEOs.

Regarding Software Engineers, in addition to a few demographic questions, the interviewer asked open-ended questions about their perception regarding the training received when they joined the company, relationships and communication among collaborators, demand for software developers, and their work autonomy. In addition, the interviewer asked the Software Engineers' opinions about obsolescence, incentives offered by the company, the lack of organizational commitment, burnout, and the salary amount offered.

Regarding CEOs, the interviewer asked questions about the training offered to newcomers in the company, the updating of new technologies, incentives offered, strategies used to retain professionals, and autonomy at work. Moreover, the interviewer asked about CEOs' perceptions regarding communication and relationships between employees and managers, employees compromising with the company, working overtime, burnout, and salary amount. Finally, we thanked the interviewees and debriefed them by informing them about what we planned to do with the data. The protocol was designed by two researchers over four months.

**Participants.** We selected the Software Engineers and CEOs to be interviewed based on convenience sampling. We invited the respondents by email, who provided detailed information about the research. All of them agreed to be interviewed. However, as mentioned previously, we first conducted a pilot interview with another Software Engineer and CEO (not included in the study) as a pretest [37]. Tables 1 and 2 show a summary of the interview participants. We assigned identifiers P2-P12 to Software Engineers, totaling 11 respondents with age ranges from 20 to 39 years, and the identifiers C2-C9 to CEOs, totaling 8 respondents aged 30 to 54 years. The Software Engineers came from different companies in Brazil, whereas the CEOs came from companies across two countries (Brazil 7 and USA 1). We conducted most of the interviews in Portuguese (Brazil) and later translated them into English.

**Transcription.** Once the data obtained from interviews are qualitative, we manually transcribed them to avoid losing any relevant information for the study. At the end of each interview, the interviewer listened to the entire interview and typed it into a text editor by differentiating the speech of the interviewee from his talk using a code, *i.e.,* each participant received identification in the coding process.

**Coding.** Most software engineering studies employ quantitative and qualitative methods to analyze the extracted data. Qualitative data can be expressed as words or pictures, while quantitative data are represented as numbers and other discrete categories [55]. One commonly used strategy to combine these methods is to extract values for quantitative variables from qualitative data, such as data collected from interviews or observations, to perform some sort of qualitative or statistical analysis. This process is called coding [55]. Since the first part of our study is interview-based, coding is the most appropriate analysis method.

---

[2]https://meet.google.com

Table 1. Demographic Information of Software Engineers

| Id | Age | Role | Country |
|----|-----|------|---------|
| P2 | 32 | Software Engineer | Brazil |
| P3 | 29 | Product Head | Brazil |
| P4 | 30 | Software Engineer | Brazil |
| P5 | 29 | Software Engineer | Brazil |
| P6 | 29 | Software Engineer | Brazil |
| P7 | 30 | Consultant | Brazil |
| P8 | 35 | Software Engineer | Brazil |
| P9 | 39 | Software Engineer | Brazil |
| P10 | 25 | Software Engineer | Brazil |
| P11 | 34 | Software Engineer | Brazil |
| P12 | 20 | Software Engineer | Brazil |

Table 2. Demographic Information of CEOs

| Id | Age | Role | Country |
|----|-----|------|---------|
| C2 | 54 | CEO | Brazil |
| C3 | 53 | CEO | Brazil |
| C4 | 30 | CEO | USA |
| C5 | 44 | CEO | Brazil |
| C6 | 35 | CEO | Brazil |
| C7 | 45 | CEO | Brazil |
| C8 | 47 | CEO | Brazil |
| C9 | 40 | CEO | Brazil |

**Data analysis.** For data analysis, we used Qualitative Content Analysis (QCA). It is a method for systematically describing the meaning of qualitative material. Such a method takes down a spiral path by arriving at an even more comprehensive sense of data at each step. The process is done by assigning successive parts of the material to categories of the coding frames [54]. With QCA, the research questions specify the perspective of how to examine the data. If other important aspects are achieved during the analysis, the contribution structure can be changed to include them as well.

Regardless of the material and research questions defined in the study, the QCA method always involves the same sequence of steps: *(i)* decision about research questions; *(ii)* selection of the material to be used; *(iii)* construction of a coding frame that usually comprises several main categories, each with its own set of subcategories; *(iv)* division of material into coding units; *(v)* test the coding framework by double-coding, followed by discussion of the units that were coded differently; *(vi)* evaluation of the coding framework in terms of coding consistency and in terms of validity and revision accordingly; *vii* encoding all materials by using the revised version of coding framework and transforming information at the case level; *viii* interpreting and presenting of findings [54].

Based on qualitative data analysis and QCA, we analyzed the interviews in four stages:

- **Familiarization with the data:** In this stage, we read the interview transcripts several times for familiarization with certain terms. When necessary, we also checked the literature for a better understanding of such terms.

- **Initial coding:** At this stage, we coded all interview transcripts, as follows: *(i)* for software engineers, we considered three topics such as **main causes for turnover**, **main reasons for not leaving the company,** and **best practices to minimize turnover**;*(ii)* whereas for the CEOs, we considered two topics such as **the main strategies used by the companies** and the **main reasons why software engineers leave the company**. After that, we labeled the interview excerpts by highlighting the meanings that had appropriate perceptions. We used QDA Mine Lite [3] to facilitate the creation of categories automatically, but we chose to perform the whole process manually for a better interpretation of the data. Indeed, the codes were identified and refined throughout the analysis.
- **Categories:** At this stage, we already had an initial list of codes. The next step was to identify similar codes in the data. We classified codes with similar characteristics into broader categories. In addition, we also had to refine the categories found.
- **Refinement:** We had a significant set of categories at this stage. The next step was to look for evidence to support the categories. It was also necessary to rename some categories for clarity in certain parts.
- **Finalization:** The final categories are available in the Section 5. For better quality in the data analysis, we conducted pair work, followed by conflict resolution meetings. The meetings always had the presence of a co-author to mediate the discussions.

### 4.2 Survey

**Protocol.** Based on the results from the literature review and interviews, we designed a 20-minute survey to further build our understanding of the factors that cause turnover and strategies used to mitigate it. The initial draft of the survey was composed of 16 questions, 14 of which were closed questions, and 2 of which were open questions. For the design of the survey, we followed Kitchenham and Pfleeger's guidelines for personal opinion surveys [32]. As one of the guidelines, previous surveys related to turnover were consulted (Section 2).

We piloted our survey with two Software Engineers to get feedback on the formulation of the questions, difficulties faced in answering the survey overall, and time to finish it. As these pilot respondents were experts in the area, we also wanted to know whether they felt we were asking the right kinds of questions or should be changing the approach. In response to their feedback, we modified the survey several times, rephrasing some questions and removing others to make it easier to understand and answer. The final version of the survey consisted of 17 questions (including demographics and consent). The pilot survey responses were used solely to improve the questions, and these responses were not included in the final results. We kept the survey anonymous, but in the end, the respondents could share their email to receive a study summary. The survey instrument is included in the supplementary material.

**Participants.** We followed a two-step approach to recruit respondents to the survey. In the first step, we sent invitations to groups of Software Engineers on social networks *(LinkedIn, Instagram, Twitter/X)* in Brazil from May 2023 to July 2023. We obtained 263 responses, of which 247 were considered valid, and 16 were disqualified for not agreeing with the filter question. In the second step, from July 2023 to November 2023, we recruited respondents abroad; one of the authors contacted potential respondents, asking them to share with other candidate respondents (snowballing). Because of this process, we could not track the total invitations. We obtained 83 responses, of which four were disqualified for not agreeing with the filter question. This led to 79 valid responses that were considered, where the survey respondents answered all questions.

---

[3]https://provalisresearch.com/products/qualitative-data-analysis-software/freeware/

The respondents were spread out over twenty-five countries and four continents. The top three countries where the respondents came from were Brazil, the United States, and Canada. The professional experience of the 326 respondents working as Software Engineers varied from one year (4.60%) to more than ten years (39.26%). Most respondents had an advanced degree (44.79%), i.e., a Master's or Ph.D., 33.12% of the respondents had a Bachelor's degree, and 7.06% graduated from high school without completing college.

**Data analysis.** We collected the ratings that our respondents provided for each closed question and converted these ratings to Likert scores from 1 (Strongly Disagree) to 5 (Strongly Agree). We computed the average Likert score of each statement related to different perspectives (e.g., reasons that can cause turnover, strategies applied by companies to minimize turnover, and statements on personal characteristics) and plotted Likert scale graphs. In addition, we used the matrix technique of Kasunic to perform data analysis [31]. The statistical analysis of the survey data examines response patterns, frequencies of different responses, which response occurred most frequently for each question, and variation in responses within a group. To present the frequency distribution, we used graphs to facilitate the visualization of the data more dynamically and clearly because it is a large amount of data.

## 5 RESULTS

This section presents the results for each of the two research questions identified in Section 1.

### 5.1 Reasons for Software Engineer's Turnover (RQ1)

This research question investigates the main reasons for the turnover of Software Engineers in the software development industry. From a literature review and interviews, we identified 19 different reasons for Software Engineers' turnover. Next, we used the survey responses to rank these 19 reasons, with Figure 2 showing the results. Software Engineers expressed their level of agreement regarding perceptions of turnover using a 5-point Likert scale: *(1: strongly disagree; 2: somewhat disagree; 3: neutral; 4: somewhat agree, 5: strongly agree)*. The top five reasons for Software Engineers' turnover are as follows: Toxic management (average Likert score for this statement is 94,17%, i.e., between "somewhat agree" and "strongly agree"), a higher salary proposal (average Likert score for this statement is 93,25%, i.e., between "somewhat agree" and "strongly agree"), lack of professional recognition (average Likert score for this statement is 83,44%, i.e., between "somewhat agree" and "strongly agree"), not having a good relationship with co-workers (average Likert score for this statement is 82,82%, i.e., between "somewhat agree" and "strongly agree"), and a desire to live new professional experiences (average Likert score for this statement is 79,75%, i.e., between "somewhat agree" and "strongly agree").

Table 3 presents an overview of the results of *RQ1* with the number of Software Engineers that identified each reason. Considering the total of respondents, 307 agreed (between "*somewhat agree*" and "*strongly agree*") with the topic *"A toxic management"* being the main reason for software engineers' turnover, which is about 94,17%. The answer of respondent **(SE278)** is a warning about the problem suffered by Software Engineers (SE). In addition, a Software Engineer who was interviewed (E) also highlighted this aspect in a comment:

> ✗ **SE278**: *"Not having a good relationship with the manager".*
> ✗ **E4**: *"I believe that toxic management, managers who do not empathize with their employees is also an exit factor. This is partly true. I think if you have a body of bosses there who are not flexible and don't care about the employees, that's a factor. Management needs to empathize with employees".*

Fig. 2. Nineteen Reasons for Software Engineers' Turnover

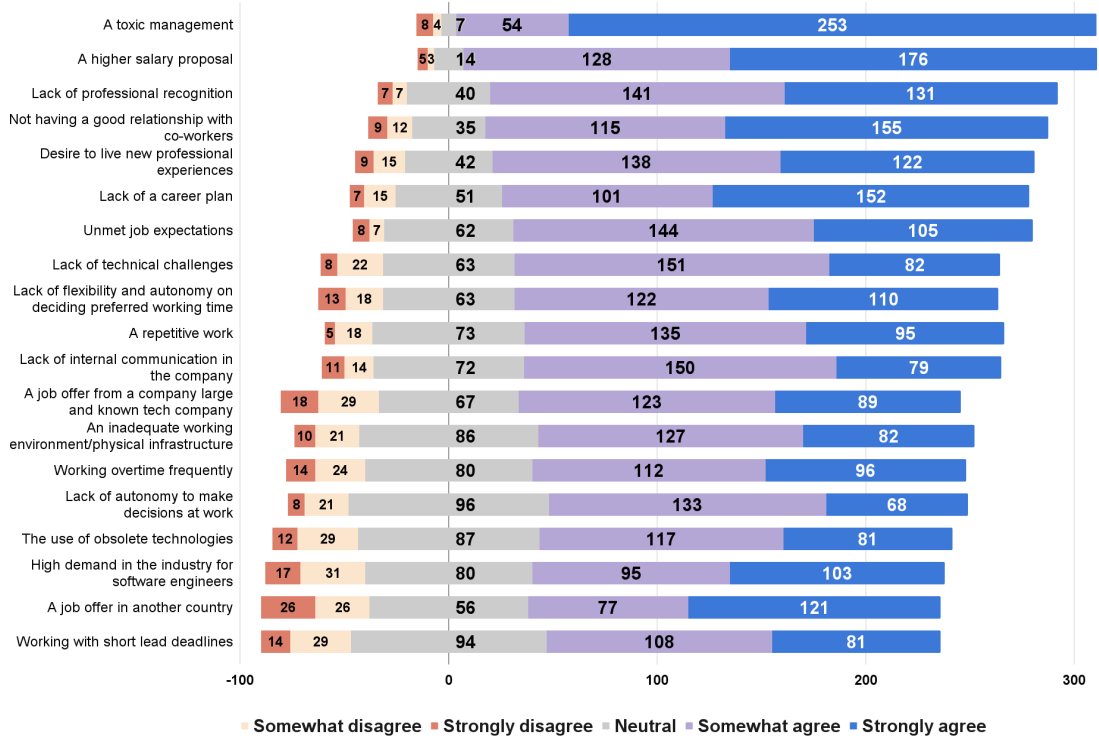

Table 3. Reasons for Software Engineers' Turnover

| **Reasons for Software Engineers' Turnover** | **Frequency** |
|---|---|
| A toxic management | 307 |
| A higher salary proposal | 304 |
| Lack of professional recognition | 272 |
| Not having a good relationship with co-workers | 270 |
| Desire to live new professional experiences | 260 |
| Lack of a career plan | 253 |
| Unmet job expectations | 249 |
| Lack of technical challenges | 233 |
| Lack of flexibility and autonomy in deciding preferred working time | 232 |
| A repetitive work | 230 |
| Lack of internal communication in the company | 229 |
| A job offer from a company large and known tech company | 212 |
| An inadequate working environment/physical infrastructure | 209 |
| Working overtime frequently | 208 |
| Lack of autonomy to make decisions at work | 201 |
| The use of obsolete technologies | 198 |
| High demand in the industry for software engineers | 198 |
| A job offer in another country | 198 |
| Working with short lead deadlines | 189 |

Shortly after, we can note that 93.25% of respondents agreed (between *"somewhat agree"* and *"strongly agree"*) that *"A higher salary proposal"* offered by competing companies tends to influence software engineers' turnover, in particular, when the payment of the wages is based on foreign solid currency. The following are comments from some of the Software Engineers that highlight such a reason:

- ✓ **SE91**: *"Occasional job changes are common. All Silicon Valley companies that pay super salaries will require overtime and short deadlines. Turnover among these companies (where the working conditions will be similar) is high because the companies' compensation system favors turnover. Another factor is the market value/package of shares offered in the company change. It is common in Silicon Valley for a software engineer to have compensation close to/upwards of $1 million because the stock has gone up 100-500-1000% since the signing of the bonus contract. This also justifies the Silicon Valley real estate bubble. Stocks, especially unicorn stocks".*
- ✓ **SE152** *"Receive in strong foreign currency".*
- ✓ **E9**: *"I think it is determined that the salary factor is the main permanence of the software developer in the organization".*
- ✓ **E10**: *"The salary, because even companies that demand more, companies with bad organizational climate, but that pay the professionals very well, and even, I worked in a company like that, and I decided to stay because of the salary".*

From the total of respondents, 83,44% of them agreed (between *"somewhat agree"* and *"strongly agree"*) that *"Lack of professional recognition"* also tends to influence Software Engineers' turnover, in particular, there is a trade-off between the professional recognition and salary proposal. The first reason can be taken into more account than the second one by some Software Engineers. The following are comments from some of the Software Engineers that highlight such theses aspects:

- ✓ **E8**: *"...I received an offer last week, where I would receive much more, but the determining factor for my staying in the company was exactly the recognition as a professional, the interpersonal relationship with the team, and all of this".*
- ✓ **SE252**: *"Salary and recognition".*

We also note that *"Not having a good relationship with co-workers"* is an important reason that influences Software Engineers' turnover in the development industry. It was considered by 82,82% of the respondents. Following are some comments from the Software Engineers interviewed that exemplify such an aspect:

- ✓ **E2** *"So, I think the issue of relationships between employees; it, is, one of the things for me, let's say, is sixty percent of maintenance, continuity in a company".*
- ✓ **E4**: *"I think it is also a super important point for the permanence of the professional. When you have a good relationship, you create bonds, and the environment becomes lighter; it is not so focused only on deliveries. So, it is also guided by how much you can motivate your team, how empathetic you can be with the pain of your colleagues, for example".*
- ✗ **E5**: *"I had only one problem with the leadership, he ended up joining the company, he didn't know the company's policy, he didn't have a vision of what the company was like, and in a way, he discouraged the team, and the team left on their own of this relationship problem".*

The reason *"Desire to live new professional experiences"* was identified by 79,75% of the Software Engineers as an important reason influencing turnover in the development industry. The following is a comment from a Software Engineer that exemplifies how to achieve new professional experiences:

- ✓ **SE67**: *"Opportunity to live in another country".*

As a survey can characterize a large group of people's knowledge, attitudes, and behaviors by studying a subset of them [31], we also evaluated whether the Software Engineers' answers were differently spread based on personal demographic factors such as *gender, generation, level of experience*, and *type of personalities*. Table 4 shows such factors spread on categories and number of respondents.

Table 4. Personal demographic factors considered in the analysis

| Factor | Category | Number of respondents |
|---|---|---|
| **Gender** | Male | = 287 |
| | Female | = 36 |
| | No-Binary | = 1 |
| | Prefer not to inform | = 1 |
| **Generation** | X (18-25) | = 44 |
| | Y (26-41) | = 221 |
| | Z (42-57) | = 56 |
| | Baby boomers (Over-57) | = 4 |
| | Prefer not to inform | = 1 |
| **Level of experience** | Less than 1 year | = 15 |
| | Between 1 and 2 years | = 37 |
| | Between 3 and 5 years | = 71 |
| | Between 6 and 8 years | = 53 |
| | Between 9 and 10 years | = 22 |
| | More than 10 years | = 128 |
| **Type of personality** | Open to new experiences | = 63 |
| | Conscientiousness | = 99 |
| | Extroversion | = 66 |
| | Kindness | = 99 |
| | Neuroticism | = 11 |

**Gender** - In our study, it was considered the genders male, female, and non-binary. Non-binary can be understood as the non-identification of the subject as a man or a woman [48]. Of the 326 respondents, 88% were male, whereas only 11% were female, and 1% was classified as non-binary.

Based on the survey data, we found that the main reasons for turnover regarding the male gender are: a higher salary proposal (94,08%); toxic management (94,08%); lack of professional recognition (82,58%); not having a good relationship with co-workers (82,23%) and; the desire to live new professional experiences (80,84%). On the other hand, the female respondents were distributed as follows: toxic management (97,22%); lack of professional recognition (94,22%); a higher salary proposal (88,89%); not having a good relationship with co-workers and (88,89%) and; lack of internal communication in the company (86,11%).

> **Observation 1:** Considering the gender perspectives, males and females agree that *a toxic management, a higher salary proposal, lack of professional recognition*, and *not having a good relationship with co-workers* are the top reasons for Software Engineers' turnover.

**Generation** - Literature uses different names for classifying the categories of the generations, and their time-categorization is sometimes differently defined. However, it does not influence the essential characteristics [4]. According to Bencsik *et al.*[4], generations can be classified as follows: *(i) Z Generation (1995 - 2010)* - it has the characteristics of net generation due to the highly developed digital era, which they were born into[61]; *(ii) Y generation (1980 - 1995)* - it is characterized by multitasking, the multi-sided, and shared attention [53]; *(iii) X generation (1960 - 1980)* - they are described as fiercely independent and en- entrepreneurial [66] and; *(iv) Baby boomers generation (1946 - 1960)* - pleasure-seeking and highly achievement-oriented [66]. Analyzing our survey data, we found that most of the respondents were between 26 and 41 years old (68% inserted into the Y Generation); 17% were between 42 and 57 years old (Z Generation); 14% were between 18 and 25 years old (X Generation); and, 1% were over 57 years old (Baby Boomers).

Regarding X Generation (18-25 years), the top five reasons for Software Engineers' turnover are as follows: toxic management (97,73%); not having a good relationship with co-workers (93,18%); a higher salary proposal (90,91%); lack of professional recognition and; lack of technical challenges (86,36%). The Y Generation (26-41 years) was distributed as: a toxic management (94,57%); a higher salary proposal (93,67%); lack of professional recognition (83,26%); not having a good relationship with co-workers (82,35%) and; desire to live new professional experiences (78,73%). According to Z Generation (42-57 years), the top five reasons for turnover were distributed as follows: a higher salary proposal (96,43%); toxic management (91,07%); desire to live new professional experiences (89,29%); lack of a career plan (87,50%) and; lack of professional recognition (85,71%). Finally, the Baby Boomers (over 57 years) considered that toxic management (100,00%), a higher salary proposal (75,00%), a desire to live new professional experiences (75,00%); repetitive work (50,00%), and not having a good relationship with co-workers (50,00%) are the main reasons for turnover.

> **Observation 2:** Generations X, Y, Z, and Baby Boomers agree that *toxic management* and *a higher salary proposal* are the top reasons for Software Engineers' turnover.

**Level of Experience** - According to [57], the longer someone spends programming, the more source code they have implemented, and the greater their experience. We classified the respondents' experience level pursuing the profession of Software Engineers as follows: *(less than 1 year)*; *(1 to 2 years)*; *(3 to 5 years)*; *(6 to 8 years)*; *(9 to 10 years)* and *(more than 10 years)*. Of the total of respondents, 39% had more than ten years of experience; 22% had between 3 and 4 years of experience; 16% had between 6 and 8 years of experience; 11% had between 1 and 2 years of experience; 7% had between 9 and 10 years of experience; and 5% had less than one year of experience.

Based on the survey data, we found that respondents with a level of experience of less than 1 year, a higher salary proposal (86,67%), lack of a career plan (86,67%), toxic management (86,67%), unmet job expectations (80,00%), and lack of internal communication in the company (80,00%) are the top five reasons for Software Engineer's turnover. The respondents with levels of experience between 1 and 2 years were distributed as follows: a higher salary proposal (97,30%); toxic management (97,30%); lack of professional recognition (94,59%); not having a good relationship with co-workers (86,49%) and; unmet job expectations (86,49%). The respondents with levels of experience between 3 and 4 years were distributed as toxic management (91,55%); a higher salary proposal (90,14%); not having a good relationship with co-workers (87,32%); lack of professional recognition (85,92%) and; lack of a career plan (83,10%). Respondents with levels of experience between 6 and 8 years considered a higher salary proposal (98,11%), toxic management (98,11%), unmet job expectations (86,79%), repetitive work (84,91%), and not having a good relationship with co-workers (83,02%) the top five reasons for Software Engineers' turnover.

Experienced professionals with levels of experience between 9 and 10 years reported that toxic management (100,00%), a higher salary proposal (95,45%), lack of a career plan (90,91%), lack of professional recognition (86,36%), and the desire to live new professional experiences (77,27%) are the main reasons for turnover. Finally, respondents with levels of experience of more than 10 years reported the following principal reasons for turnover: toxic management (92,97%); a higher salary proposal (92,19%); not having a good relationship with co-workers (83,59%); lack of professional recognition (81,25%) and; the desire to live new professional experiences (78,13%).

> **Observation 3:** Considering the experience perspective, *less than 1 year, 1 to 2 years, 3 to 5 years, 6 to 8 years, 9 to 10 years*, and *more than 10 year* agree that *a higher salary proposal* and *a toxic management* are the top reasons for Software Engineers' turnover.

**Type of personality** - The *Big Five Factor Model* [14, 38] is a widely accepted theory that classifies personality into five main dimensions: *Openness to experience* - Reflects the degree of intellectual curiosity, creativity, and a preference for novelty and variety; *Conscientiousness* - Indicates a tendency to show self-discipline, to act dutifully, and to aim for achievement; *Extroversion* - Energy, positive emotions, assertiveness, sociability, the tendency to seek for stimulation in the company of others, and talkativeness describe this trait; *Kindness* - Expresses a tendency to be compassionate and cooperative rather than suspicious and antagonistic towards others; and *Neuroticism* - Reflects the tendency to frequently experience unpleasant emotions, such as anger, anxiety, depression, or vulnerability. In our study, most of the respondents presented a personality fitted into Conscientiousness (39%) and/or Kindness (39%); 20% fitted into Extroversion; 19% fitted into Open to new experiences; and 3% fitted into Neuroticism.

Considering the first dimension (respondents with personality traits *open to new experiences*), a higher salary proposal (95,24%), toxic management (95,24%), lack of professional recognition (90,48%), unmet job expectations (87,30%), and the desire to live new professional experiences are the top five reasons for Software Engineer's turnover. Next, we examined the respondents with personality traits of *conscientiousness*, and they were distributed as follows: a higher salary proposal (96,97%); toxic management (95,96%); lack of professional recognition (89,90%); lack of a career plan (84,85%) and; the desire to live new professional experiences (83,84%). The respondents with personality traits of *extroversion* were distributed as follows: a higher salary proposal (95,45%); toxic management (89,39%); lack of professional recognition (84,85%); lack of a career plan (75,76%); and unmet job expectations (75,76%).

Finally, we analyzed the respondents with personality traits of kindness and neuroticism. The respondents with personality traits of *kindness* reported that a higher salary proposal (97,98%), toxic management (96,97%), lack of professional recognition (86,87%), desire to live new professional experiences (83,84%), and not having a good relationship with co-workers (79,80%) are the top five reasons for Software Engineer's turnover. Regarding respondents with personality traits of *neuroticism*, a higher salary proposal (100,00%), repetitive work (90,91%), toxic management (81,82%), unmet job expectations (81,82%), and lack of professional recognition (81,82%) are the top five reasons for turnover.

> **Observation 4:** Considering the personality perspectives, *open to new experiences, conscientiousness, extroversion, kindness*, and *neuroticism* agree that *a higher salary proposal*, *toxic management*, and *lack of professional recognition* are the top reasons for Software Engineers' turnover.

In general, by considering the association between two personal demographic factors, it can be possible to determine if there are nonrandom associations between two categorical variables applying the Fisher's Exact Test [17]. A small p-value (typically less than the chosen significance level, e.g., 0.05) indicates that the observed association is unlikely to occur by chance, leading to the rejection of the null hypothesis. This way, whether the p-value is less than the chosen significance level, it is possible to conclude that there is a significant association between the two categorical variables.

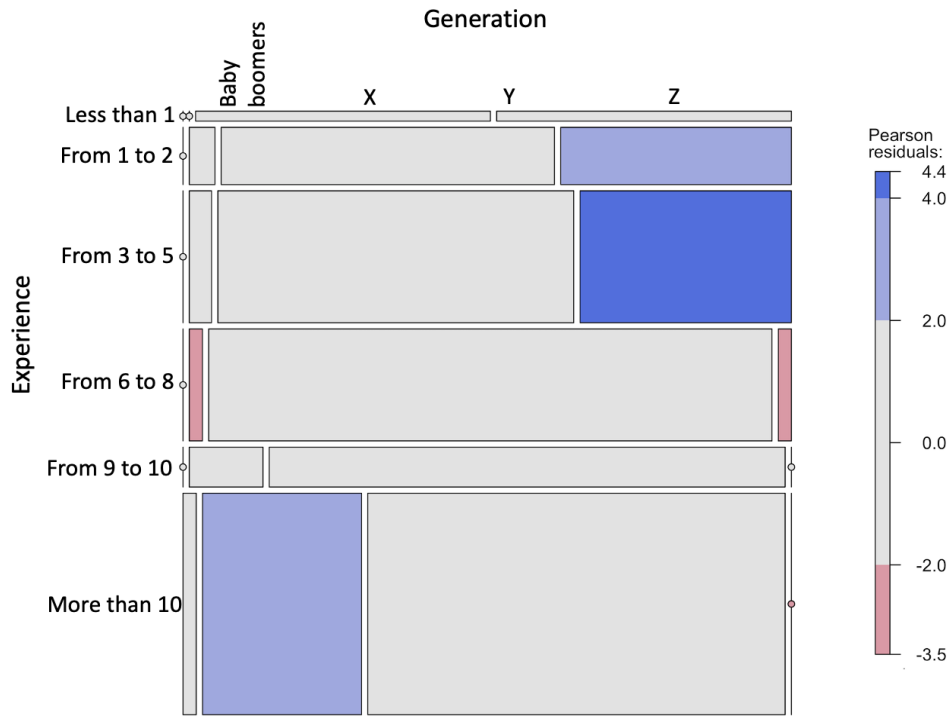

Fig. 3. Mosaic Chart showing the association between Generation and Experience.

Aiming to understand the relationship among the variables (**Gender and Generation**, **Gender** and **Level of experience**, **Gender** and **Types of personality**, **Generation** and **Level of experience**, **Generation** and **Types of personality**, and **Level of experience** and **Types of personality**) in the turnover context, we applied Fisher's exact test to determine whether they have a significant association or relationship. The results show that only the **Generation** and **Level of experience** are associated with all of the nineteen factors.

> **Observation 5:** Based on the association between the personal demographic factors, we found that *Generation* and *Level of experience* have an association for all of the nineteen reasons for Software Engineers' turnover. Considering such an association perspective, *a higher salary proposal* and *toxic management* are the top reasons for Software Engineers' turnover.

We used the mosaic chart (Figure 3) to understand how each aspect influenced the association. A mosaic chart uses Pearson residual values to assess the association or independence among two categorical variables. These residual values help to identify areas of the chart where observed and expected frequencies significantly differ, indicating a potential association among the variables. The following formula was used to compute the residual measures ($PearsonResidual = (ObservedFrequency - ExpectedFrequency)/\sqrt{(ExpectedFrequency)}$). The residual measures how far the observed frequency deviates from what would be expected if the variables were independent. A positive residual indicates that the observed frequency in a cell is higher than expected under the assumption of independence. This suggests an association or dependency between the two categorical variables. On the other hand, a negative residual indicates that the observed frequency in a cell is lower than expected under the assumption of independence. This suggests that there is less of a specific category interaction than expected. A residual close to zero suggests that the observed and expected frequencies are similar, indicating that the variables are independent for that cell. All Python scripts, results, and charts are available on the project website.

> **RQ1 Summary:** *Toxic management*, *a higher salary proposal*, *lack of professional recognition*, *not having a good relationship with co-workers*, and *a desire to live new professional experiences* are the top five reasons for Software Engineers' turnover.

## 5.2 Effective strategies that CEOs use to mitigate Software Engineer's turnover (RQ2)

From the interviews with CEOs, we identified 20 strategies to reduce Software Engineer's turnover. We asked open questions to the eight CEOs based on the step-by-step described in Section 4.1, which resulted in identifying strategies used in companies to retain Software Engineers naturally. We used the number of mentions as a criterion to quantify the strategies used by CEOs concerned with the retention of employees in companies. The following are the strategies that are more frequently mentioned by CEOs in interviews.

**Promote technological challenges** - It was also one of the most cited strategies for retention, according to CEOs **C**. According to CEO **C4**, Software Engineers are driven by challenges by promoting technological challenges and encouraging professionals to stay at the company.

> ✓ **C4**: ..." *activities that make you grow as a professional, you learn, challenge yourself, and get out of your comfort zone, I think this is the first fundamental point, I think that the vast majority of good technology professionals become demotivated when they start doing tasks repetitive and that doesn't force you to learn new things."*

The CEO **C2** reinforces that the technological challenge is up to the professionals, who will always feel challenged. Encouraging Software Engineers with technological challenges is a significant strategy that such professionals are constantly learning about.

> ✓ **C2**: ..." *and if this decreases in turnover, I think yes, no doubts, because I think the issue of turnover thinking about it more broadly, for you to retain a person, for you to engage a person in the company, you have to have a technological challenge at the height, that she always feels challenged."*

**Ensuring a good working environment** - Promoting a work environment with adequate physical structure and better technologies is a strategy that must be adopted for professionals to remain in the company. The following are some CEOs' comments where one reinforces that a stimulating work environment improves professional performance.

> ✓ **C2**: ...*that makes people feel good, an environment that is stimulating to work in, an environment where the person doesn't think: today is Sunday, tomorrow I have to go to work, on the contrary, the person, hey, tomorrow is Monday, I'm here with all my energy, full gas to go to work.*
> ✓ **C4**: ...*yeah, and finally give the person a good structure to work with, right? So, people working with the best technologies, people working with a good setup, it is, in the context of work, in the physical environment, office, having a good environment, for people to feel comfortable."*

We observe that CEOs are increasingly focused on promoting an adequate and comfortable work environment both physically and technologically. As a result, it is possible that the professionals feel pleasure in carrying out their activities and increasing productivity by avoiding turnover.

**Training program** - Training programs attract young people to be trained according to the company's needs and can contribute to the professionals' stay in their jobs for longer. Education and training program is a strategy used by C8 to reduce turnover as follows:

> ✓ **C8**: ..." *So, our turnover is largely being resolved through a training program for practical people within the company, attracting increasingly younger and less experienced people so that we can draw a workforce pipeline that guarantees us sustainability. Then, in the teams, we always keep one or two key guys who are more experienced."*

**Offer adequate remuneration** - In general, aligning the salary with the market or paying an above-average salary has been used as a standard strategy within companies. CEO **C4** commented that having a base salary that is compatible with the market is a strategy used by his company. In addition, interviewee **C5** also reinforces this aspect.

> ✓ **C4**: ..." *which is always having a good salary table, above the market average, this is something we have always done, in addition to good salaries, having this type of policy, for example, better employees have access to the opportunity to become a partner, become a shareholder, etc ... ".*
> ✓ **C5**: ..." *we offer remuneration that is very competitive in the market. We are not the company that pays the most, but we also do not pay below the market. I think we are very competitive there."*

**Be open-minded to new generation** - We can observe that the new generation desires to learn and feel a part of the organizational environment. In this sense, the company needs to promote such aspects to retain professionals in the software industry. Interviewee **C7** stands up for honest relationships to show the new generation the possibility of engagement and growth in the company.

> ✓ **C7**: ..."*You have a younger generation, generation Z mainly, that comes with this expectation of fast growth and career projection, and then we are looking more effectively to make these people feel part of it. So, it certainly helps to have the notion that they are important within the structure and are heard, okay?".*

**Offer mentoring during onboarding** - The recruitment and selection process is a key process for retention. This becomes more robust when mentored by an experienced professional. Interviewee **C2** affirms the importance of assigning a mentor to a new professional.

> ✓ **C2**: ..." *any new employee who joins the company has a sponsor who is formally designated for his or her first three months at the company, and it has been very effective."*

**Promote project-specific training** - This aspect is related to the previous strategy, and we could observe the need for employee training in specific projects with the collaboration of a leader. CEO **C6** highlights the importance of training when changes occur in the project.

> ✓ **C6**: ..."*Yes, when there is, yes, we do training; when there is a change, we do training or more than training, usually, we bring a leader, or a person who specializes in the subject, to teach, you know? Doing something."*

> **Observation 6:** *Promoting technological challenges, ensuring a good working environment, training programs, offering adequate remuneration,* and *be open-minded to new generation* are the top five strategies used by CEOs to reduce turnover in the software industry.

Although we have collected strategies to reduce software engineers' turnover from the perspective of CEOs, this research question investigates the most effective strategies that CEOs can use to mitigate turnover from the point of view of software engineers. The survey responses identified the 18 most effective strategies for CEOs in the software development industry. In general, all strategies CEOs mentioned during interviews complement those that Software Engineers indicated in the survey. Figure 4 shows an overview of the results based on the answers from the Software Engineers.

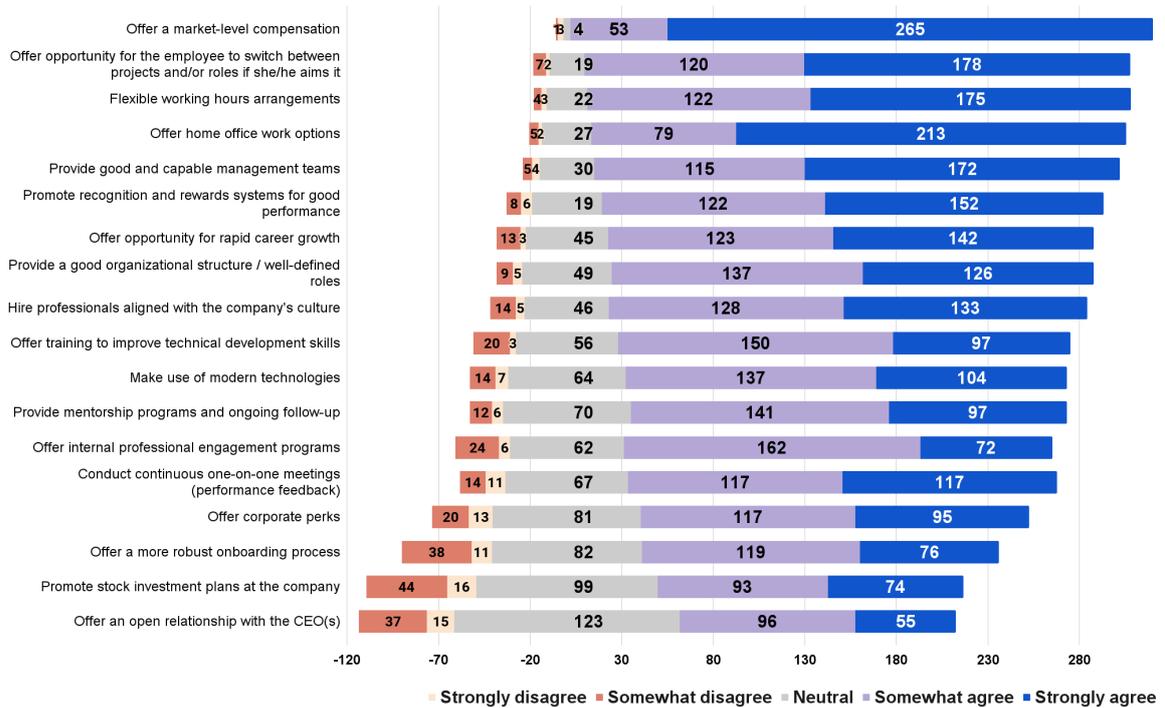

Fig. 4. Most effective strategies to mitigate Software Engineers' Turnover

Software Engineers expressed their level of agreement regarding the most effective strategies to mitigate turnover using a 5-point Likert scale (**1**: strongly disagree; **2**: somewhat disagree; **3**: neutral; **4**: somewhat agree, **5**: strongly agree). The top five more effective strategies used by CEOs to mitigate Software Engineers' turnover are as follows: offer a market-based pay structure (average Likert score for this statement is 97,55%, i.e., between "somewhat agree" and "strongly agree"); offer an opportunity for the employees to switch between projects and/or roles if they aim it (average Likert score for this statement is 91,41%, i.e., between "somewhat agree" and "strongly agree"); flexible working hours arrangements (average Likert score for this statement is 91,10%, i.e., between "somewhat agree" and "strongly agree"); offer home office work options (average Likert score for this statement is 89,57%, i.e., between "somewhat agree" and "strongly agree") and; provide good and capable management teams (average Likert score for this statement is 88,04%, i.e., between "somewhat agree" and "strongly agree").

Table 5 presents an overview of the results of *RQ2* with the number of Software Engineers that identified each strategy. Considering the 326 respondents, 318 stated that *"Offer a market-based pay structure"* is the most effective strategy CEOs use to mitigate Software Engineers' turnover (97,55%). However, Software Engineers primarily address aspects related to motivation and satisfaction with the job. The following are some comments from interviewees that highlight these aspects:

- ✓ **C4**: *"...our philosophy regarding the subject is that the salary has to be high enough for a person not to worry about the salary".*
- ✓ **SE32**: *"I believe that one of the important factors to consider for a deeper analysis of turnover in SE may be weakly related not to the environment that the person is inserted but, on the other hand, strongly related to the opportunities that people have to feel happier in what they are doing".*

Table 5. Most effective strategies used by CEOs to mitigate Software Engineers' Turnover

| Strategy | Frequency |
|---|---|
| Offer a market-based pay structure | 318 |
| Offer opportunity for the employees to switch between projects and/or roles if they aim it | 298 |
| Flexible working hours arrangements | 297 |
| Offer home office work options | 292 |
| Provide good and capable management teams | 287 |
| Offer opportunity for rapid career growth | 265 |
| Provide a good organizational structure / well-defined roles | 263 |
| Offer an open relationship with the CEO(s) | 151 |
| Provide mentorship programs and ongoing follow-up | 238 |
| Promote recognition and rewards systems for good performance | 274 |
| Hire professionals aligned with the company's culture | 261 |
| Offer training to improve technical development skills | 247 |
| Make use of modern technologies | 241 |
| Conduct continuous one-on-one meetings (performance feedback) | 234 |
| Offer internal professional engagement programs | 234 |
| Offer corporate perks | 212 |
| Offer a more robust onboarding process | 195 |
| Promote stock investment plans at the company | 167 |

Shortly after, we identified that 91,41% of respondents agreed (between "somewhat agree" and "strongly agree") that "Offer an opportunity for the employees to switch between projects and/or roles if they aim it" tends to reduce turnover intentions on the part of Software Engineers. Following are some comments from the interviewees reinforcing the importance of switching from one project to another when the employee wants to work with new technologies:

- ✓ **C3**: "...So, if I have a professional on project A, he is not happy; he wants to work on another technology, and there is project B that is an opportunity for him; we insist on the idea of moving, obviously in a planned way, right? So, the demands resulting from the employee evaluation journey that generates movement at the Company."
- ✓ **SE273**: "...ask whether someone is self-employed, where switching projects/companies comes more naturally.".

> **Observation 7:** According to our survey data with software engineers, the top five more effective strategies used by CEOs to mitigate Software Engineers' turnover are as follows: *offer a market-based pay structure*; *offer an opportunity for the employee to switch between projects and/or roles if they aim it*; *flexible working hours arrangements*; *offer home office work options* and; *provide good and capable management teams*.

After the general analysis, we investigated whether the Software Engineers' responses were differently spread based on personal demographic factors, such as *gender*, *generation*, *level of experience*, and *type of personality*.

**Gender** - Regarding the male gender responses, the most effective strategies were distributed as follows: offer a market-based pay structure (97,56%), offer an opportunity for the employee to switch between projects and/or roles if she/he aims it (90,94%), flexible working hours arrangements (90,94%), offer home office work options (89,90%), and provide good and capable management teams (87,11%). On the other hand, the female responses were distributed as follows: offer a market-based pay structure (97,22%), offer an opportunity for the employee to switch between projects and/or roles if she/he aims it (94,44%), provide good and capable management teams (94,44%), hire professionals aligned with the company's culture (94,44%), and offer training to improve technical development skills (91,67%).

> **Observation 8:** Considering the gender perspectives, males and females agree that *offer a market-based pay structure*, *offer opportunity for the employee to switch between projects and/or roles if they aim it*, and *provide good and capable management teams* are the top strategies used by CEOs to mitigate Software Engineers' turnover.

**Generation** - Regarding X Generation (18-25 years), the top five strategies used by CEOs to mitigate Software Engineers' turnover are as: offer a market-based pay structure (84,09%), provide good and capable management teams (79,55%), offer opportunity for the employee to switch between projects and/or roles if they aim it (75,00%), offer home office work options (77,27%), and flexible working hours arrangements (75,00%). According to Y Generation (26-41 years), the results were also similar: offer a market-based pay structure (86,88%), flexible working hours arrangements (81,45%), offer an opportunity for the employee to switch between projects and/or roles if they aim it (81,00%), offer home office work options (79,64%), and provide good and capable management teams (77,38%).

Finally, Generation Z and Baby Boomers chose different strategies to mitigate turnover. Regarding Z Generation (42-57 years), important aspects are related to offering a market-based pay structure (87,50%), offering an opportunity for the employee to switch between projects and/or roles if they aim it (83,93%), offering home office work options (82,14%), flexible working hours arrangements (82,14%), and offer an opportunity for rapid career growth (76,79%). According to Baby Boomers (over 57 years), offer an opportunity for the employee to switch between projects and/or roles if they aim it (75,00%), offer home office work options (75,00%), flexible working hours arrangements (75,00%), promote stock investment plans at the company (75,00%), and hire professionals aligned with the company's culture (75,00%) are the most effective ones.

> **Observation 9:** Considering the generation perspectives, X, Y, Z, and Baby Boomers agree that *Offer opportunity for the employee to switch between projects and/or roles if they aim it* and *flexible working hours arrangements* are the top strategies used by CEOs to mitigate Software Engineers' turnover.

**Level of Experience** - According to novices (less than 1 year of experience), offers a more robust onboarding process (86,67%), offers training to improve technical development skills (86,67%), offers an opportunity for rapid career growth (86,67%), provide mentorship programs and ongoing follow-up (86,67%), and offer home office work options (86,67%) are the top five strategies used by CEOs to mitigate Software Engineers' turnover. Still considering professionals at the beginning of their career (levels of experience between 1 and 2 years ), the respondents consider that offering a market-based pay structure (78,38%), offers an opportunity for the employee to switch between projects and/or roles if they aim it (72,97%), provides good and capable management teams (72,97%), provides a good organizational structure/well-defined roles (72,97%), and offers home office work options (70,27%) are the most effective ones.

Software engineers in the middle of their careers have a similar perception of turnover. Respondents with levels of experience between 3 and 5 years consider that offer a market-based pay structure (88,73%), offer an opportunity for the employee to switch between projects and/or roles if they aim it (87,32%), offer home office work options (85,92%), flexible working hours arrangements (85,92%), and offer an opportunity for rapid career growth (81,69%) are the top five strategies used by CEOs to mitigate turnover. Considering respondents with levels of experience between 6 and 8 years, offer a market-based pay structure (79,25%), offer an opportunity for the employee to switch between projects and/or roles if she/he aims it (73,58%), provide good and capable management teams (71,70%), offer home office work options (73,58%), and flexible working hours arrangements (73,58%) are the most effective ones.

Finally, experienced software engineers (levels of experience between 9 and 10 years) consider offering a market-based pay structure (86,36%), provide good and capable management teams (81,82%), provide a good organizational structure/well-defined roles (81,82%), flexible working hours arrangements (77,27%), and offer an opportunity for the employee to switch between projects and/or roles if they aim it (72,73%) are the top strategies used by CEOs to mitigate Software Engineers' turnover. Respondents with levels of experience of more than 10 years reported that offering a market-based pay structure (89,84%), flexible working hours arrangements (83,59%), offering an opportunity for the employee to switch between projects and/or roles if they aim it (82,81%), provide good and capable management teams (80,47%), and offer home office work options (81,25%) are the most effective ones.

> **Observation 10:** Considering the different software engineers' levels of experience, *offer home office work options* and *offer a market-based pay structure* are the top strategies used by CEOs to mitigate turnover.

**Types of personality** - Respondents with personality traits *open to new experiences* consider that offer a market-based pay structure (90,48%), flexible working hours arrangements (88,89%), promote recognition and rewards systems for good performance (85,71%), hire professionals aligned with the company's culture (77,78%), and offer home office work options (80,95%) are the top five strategies used by CEOs to mitigate Software Engineers' turnover. Respondents with personality traits of *conscientiousness* believe that offer a market-based pay structure (80,81%), flexible working hours arrangements (76,77%), offer home office work options (74,75%), provide good and capable management teams (72,73%), and promote recognition and rewards systems for good performance (72,73%) are the top ones.

On the other hand, respondents who were more energetic and assertive (personality traits of *extroversion*) reported that the most effective strategies offer a market-based pay structure (81,82%), promote recognition and rewards systems for good performance (75,76%), offer home office work options (77,27%), flexible working hours arrangements (77,27%), and offer an opportunity for the employee to switch between projects and/or roles if they aim it (72,73%). Respondents who express a tendency to be compassionate and cooperative (personality traits of *kindness*) believe that offering a market-based pay structure (86,87%), flexible working hours arrangements (83,84%), offer an opportunity for the employee to switch between projects and/or roles if they aim it (82,83%), offer home office work options (83,84%), and provide good and capable management teams (80,81%) are effective strategies to mitigate turnover.

Finally, respondents with personality traits of *neuroticism* reported that offer an opportunity for the employee to switch between projects and/or roles if they aim for it (90,91%), offer a market-based pay structure (90,91%), offer a more robust onboarding process (81,82%), promote recognition and rewards systems for good performance (81,82%), and offer home office work options (81,82%) are the top five strategies used by CEOs to mitigate Software Engineers' turnover.

Fig. 5. Generation and Experience association.

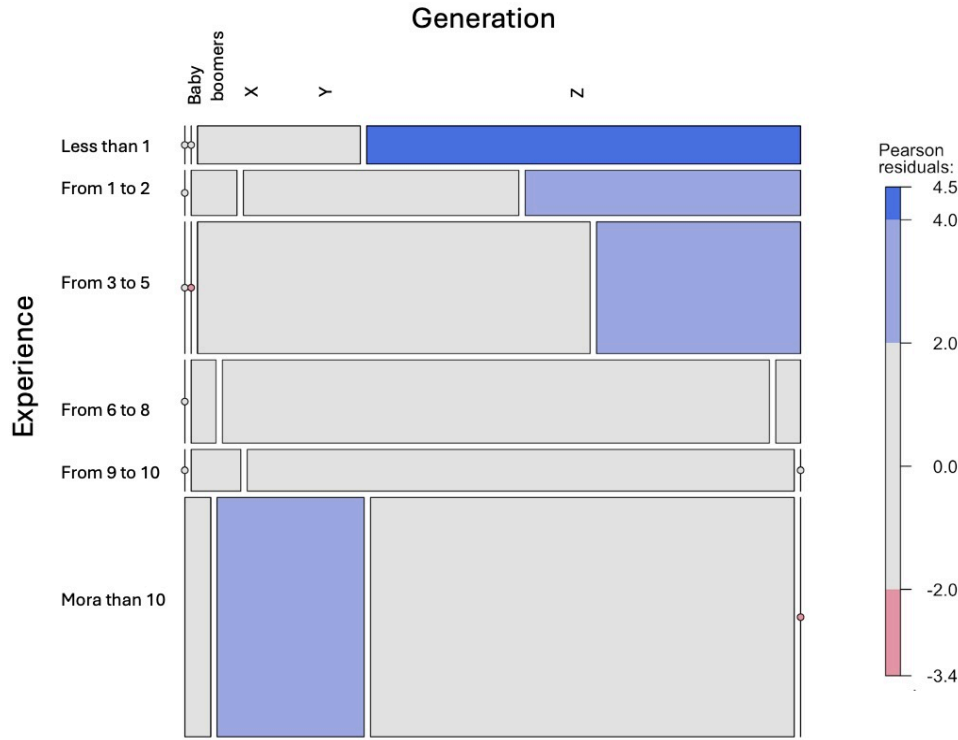

**Observation 11:** Considering the different personality perspectives, *offer a market-based pay structure* and *offer home office work options* are the top strategies used by CEOs to mitigate Software Engineers' turnover.

Aiming to understand the relationship between the different aspects (**Gender and Generation**, **Gender** and **Level of experience**, **Gender** and **Types of personality**, **Generation** and **Level of experience**, **Generation** and **Types of personality**, and **Level of experience** and **Types of personality**) in the turnover context, we applied Fisher's exact test to determine whether they have a significant association. As in RQ1, the results show that only the **Experience** and **Generation** are associated with all eighteen strategies. Figure 5 shows the mosaic chart used to understand these relationships. All Python scripts, results, and charts are available on the project website.

**Observation 12:** Based on the association between the personal demographic factors, we identified that *Generation* and *Level of experience* have an association for all of the eighteen strategies used by CEOs to mitigate Software Engineers' turnover. Based on this association, *offer the opportunity for the employee to switch between projects and/or roles if they aim for it*, *offer home office work options*, and *flexible working hours arrangements* are the top strategies to mitigate Software Engineers' turnover.

> **RQ2 Summary:** *Offer a market-based pay structure*, *offer opportunity for the employee to switch between projects and/or roles if they aim it*, *flexible working hours arrangements*, *offer home office work options*, and *provide good and capable management teams* are the top strategies used by CEOs to mitigate Software Engineers' turnover.

## 6 DISCUSSION

Our study examines why Software Engineers leave their jobs and CEOs' strategies to keep such professionals in the software development industry. The key findings presented in this study indicate that "toxic management" is perceived by Software Engineers as the primary factor contributing to turnover. Furthermore, from the point of view of Software Engineers, "market-level compensation" is the most effective retention strategy that CEOs can use to mitigate turnover. The prominence of "toxic management" as the main turnover factor underscores the critical importance of leadership and management skills in software companies. Toxic management practices often lead to a hostile work environment, negatively impacting employee morale and productivity. Thus, the companies must define a policy of good dealing with collaborators to the management became one of the reasons to keep them in their job and not the opposite as was pointed out by CEOs **C6** and **C9**:

- ✓ **C6**: *"I have a lot of people staying in the company because of the management, and I have a lot of leaving because of the management too"*.
- ✓ **C9**: *"We believe that the challenge is also a fundamental part of keeping these people motivated and happy within the company, so we have a very participatory management model, where people are very proactive in deciding how things are going to be done, right?"*

Overall, the findings suggest that improving management practices could significantly reduce turnover rates. The emphasis on "market-level compensation" highlights that competitive salaries and a positive work environment are crucial in retaining talent. Besides, the correlation between "repetitive work" and aspects of experience and generation suggests that different groups of engineers may have distinct needs and preferences. For example, more experienced engineers or those from newer generations might seek more challenging and innovative projects, indicating that retention strategies should be tailored to meet these varying needs as mentioned by CEOs **C7** and **C9**:

- ✓ **C7**: *"Now a younger generation, they are there wanting to learn, they are there engaged, they want to be part of this stand-up ecosystem, they also want to have a prediction that, because I am working with a new technology, I will not be one more in the crowd, I'll talk to the director of the company if it's possible, anyway. So I think access to several people in the company, a frank relationship"*.
- ✓ **C9**: *" We also believe that the work environment has to be very important, nice light, good people, this is in our culture, our number one value is good people, not good people, it's good people and good people it's that person you enjoy working with, so our culture is very strong in that sense"*.

The results of this study align with previous research presented by *Buhari et al.* [9], *Farooq et al.* [15], *Sharma et al.* [56], which has consistently identified the importance of a positive work environment, fair compensation, and also highlighted the significance of salary policies and professional recognition as key factors in talent retention within the software industry. However, our study extends the literature by providing additional insights into the specific practices and perceptions within the Software Engineering field. Identifying toxic management as a predominant turnover factor adds depth to our understanding, suggesting that addressing interpersonal and managerial issues is as critical as financial incentives.

For software companies, developing management strategies that foster a positive work environment and prioritize employee well-being is essential. Implementing competitive compensation policies and professional development programs can significantly reduce turnover rates. Companies should also consider training managers to mitigate toxic management practices and promote a more supportive and inclusive workplace culture. Furthermore, recognizing the diverse needs of different employee groups, such as offering more challenging projects to experienced engineers, can enhance job satisfaction and retention. Customized retention strategies that consider the specific preferences of various demographic groups can lead to more effective retention outcomes.

## 6.1 Implications for Industry

The findings presented in our study offer several practical implications for the software industry. Firstly, identifying "toxic management" as a key factor driving turnover emphasizes the critical need for companies to invest in leadership training and development programs. Organizations should focus on fostering a supportive and inclusive management culture, promoting positive leadership behaviors, and addressing any instances of toxic management swiftly and effectively [62]. Secondly, "market-level compensation" as an effective retention strategy highlights the importance of maintaining competitive salary structures. Software companies should regularly benchmark their compensation packages against industry standards to ensure they are offering attractive and fair wages. In addition to competitive pay, companies should consider implementing performance-based incentives and comprehensive benefits packages to enhance overall employee satisfaction. Thirdly, the correlation between "repetitive work" and turnover suggests that job enrichment and task variety are crucial in retaining Software Engineers. Employers should strive to provide challenging and engaging work assignments, opportunities for skill development, and clear career progression paths. Implementing job rotation programs and encouraging involvement in innovative projects can help mitigate the negative effects of repetitive work and keep employees motivated and committed.

Furthermore, the findings reinforce the importance of a positive work environment as was also reported by *Buhari et al.* [9] and *B. Johnson at al.* [28]. Companies should prioritize creating a healthy workplace culture that values employee well-being, work-life balance, and mental health. Initiatives such as flexible work arrangements, wellness programs, and opportunities for social interaction can contribute to a more satisfying and supportive work environment. Finally, with the growing prevalence of remote and hybrid work models, software companies must adapt their retention strategies to the new work landscape. This includes investing in technology and tools that facilitate effective remote work, maintaining regular and transparent communication, and ensuring remote employees feel connected and valued [51]. By addressing remote work's unique challenges and opportunities, companies can better retain their talent in a competitive industry.

## 6.2 Implications for Research

Our study's findings provide several implications in both industry and academia, which can drive future research. First, identifying "a toxic management" as a primary driver of turnover highlights the need for further investigation into managerial practices and their impact on employee retention. Future studies could delve deeper into the specific behaviors and characteristics of toxic management that most significantly affect Software Engineers, and explore interventions that can mitigate these negative effects. In addition, the strong emphasis on "market-level compensation" as an effective retention strategy suggests that financial incentives remain crucial. However, it also opens up avenues for research into non-financial incentives, as shown in Table 5 and their relative impact on different demographic groups within the software industry. Understanding how different generations, experience levels, and personal preferences interact with various retention strategies can provide a more comprehensive view of what drives software engineers to stay with or leave their organizations.

Third, the correlation between "repetitive work" and turnover underscores the importance of defining the job design and task variety to retain talent. Future research could examine how different work assignments and project rotations influence job satisfaction and turnover rate. In addition, exploring the role of manager and work team while facilitators of career development and continuous learning to enhance job satisfaction could yield valuable insights. Lastly, considering the evolving nature of work, especially with the rise of remote and hybrid work models post-pandemic [7], future research should investigate how these new work arrangements affect turnover in the software industry. Studies could assess the long-term implications of remote work on employee engagement, job satisfaction, and organizational loyalty, providing a modern perspective on retention strategies in a digital-first work environment.

## 7 THREATS TO VALIDITY

In this section, we discuss several threats to the validity of our study.

### 7.1 Conclusion validity

Threats to conclusion validity are concerned with issues that relate to the treatment and the outcomes of the study, including, for instance, the choice of sample size and, as another example, the care taken in implementing a survey [64]. We conducted interviews with open-ended questions in which the participants were asked to provide their perceptions and points of view. The interviews were then corroborated through a survey. The interviews were conducted at different companies, and when they happened within the same company, the participants were warned not to talk to each other about it to avoid bias. We approached the design and implementation of the survey with the same level of care.

The quality of the material used in the study is also critical for the conclusion's validity. For this reason, we conducted a pilot interview with a software engineer and a CEO, besides applying a survey pre-testing with two software engineers to ensure that the interview prompt and survey instrument were high quality. In addition, we carefully validated the interviews and findings with the participants as we performed the analysis. Indeed, we asked for clarification when so needed to avoid the threat of drawing false conclusions based on the interview data.

### 7.2 Internal validity

We kept our questions open-ended and let participants talk most of the time to reduce introducing interview bias during the interviews. Moreover, we concluded the interviews by asking the participants whether they had further thoughts and gave them ample time to respond before concluding the interviews. This is because the participants might not have mentioned some points they could have brought up, given more time to think. In the same way, we asked survey participants to answer a question about any additional thoughts they wanted to share before concluding the survey.

We also attempted to reduce the bias of conducting remote participant interviews. Compared with interviews in person, the interviews remotely performed might introduce some bias, such as incomplete, shorter, or unclear answers. Thus, we took care that the interviewees gave us more clarification when needed to understand their answers retrospectively. As the survey questions were derived from the answers from the interviews, it is possible that the questions on the survey might not have been sufficiently representative (e.g., additional aspects that induce Software Engineers' turnover and effective strategies to reduce it from the point of view of both, CEOs and Software Engineer). It was mitigated by allowing participants to provide additional thoughts through an open-ended question at the end of the survey.

## 7.3 Construct validity

There are potential threats to construct validity from the lack of a clear definition of a software engineer and CEO. Participants generally understood that, in the first case, we meant someone in the development team responsible for coding activities. On the other hand, CEOs deal with strategic decisions and direct the company's overall growth. In addition, we verbally clarified whenever there appeared to be some confusion, both at the start of the interviews and throughout.

Another threat to construct validity is related to the potential problem of evaluation apprehension [64]. It was mitigated by letting the participants know they would remain anonymous as well as by assuring them that all information gathered during the interviews and survey would solely be used only by the research team and never shared beyond.

## 7.4 External validity

Our 19 interviews were conducted with Software Engineers (11) and CEOs (8) working in 19 different companies. Even though these interviews yielded relevant insights, they can be considered a small sample. Moreover, we only sampled software engineers and CEOs from Brazil, and the findings may not be generalized to other countries and companies.

The same threat exists concerning the participants in the survey. Even though the respondents reside in twenty-five countries across four continents, our findings may not be generalized to represent the experiences and perceptions of all Software Engineers.

## 8 CONCLUSION

Turnover is a phenomenon that has significantly affected the software industry since intellectual capital is important for the progress of projects. Despite the efforts made by researchers and professionals to minimize the turnover, more studies are needed to understand critical factors to avoid the loss of talented professionals. Therefore, our study contributes to the understanding of factors influencing turnover among software engineers and offers practical insights for implementing effective retention strategies. We used a mixed approach, with qualitative and quantitative studies on Software Engineers' perception of turnover and more efficient strategies used in the software development industry aiming to reduce it. By addressing organizational and personal factors, software companies can develop more holistic and effective approaches to retain talent. Future research should continue to expand on these findings, exploring new dimensions and contexts to enrich our understanding of turnover in the software engineering industry.

We investigated Software Engineers' perceptions of turnover (RQ1), and the resulting findings showed that having toxic management in a project is the factor that hurts the retention of such professionals. Conversely, salary plays a significant role in retention. We also investigated which strategies companies use are most efficient for retaining professionals in their jobs (RQ2). The findings showed that remuneration is the main attraction for employment. Moreover, the project manager has a fundamental role in promoting opportunities for changes between projects which collaborates in retaining these professionals. The findings also showed that companies are promoting a comfortable working environment and keeping Software Engineers increasingly challenged with new technologies and projects.

Additionally, we assessed personal demographic data to measure the correlation of factors causing turnover. The findings showed a positive significance regarding repetitive work when considering a possible correlation between experience levels and generations. This means that a professional's stay on the same project for a long time can reduce turnover intention. In addition, this study's practical and social implications contribute to creating more efficient strategies for retaining Software Engineers in the company. In general, the findings of our study can be useful for project managers in making decisions and retaining their most qualified professionals, and consequently, the annual loss will be reduced, and recruitment and selection, training costs, and productivity will be increased. Therefore, senior management should focus on strategies to improve their work environment. Indeed, our study can contribute as an initial guide to determine the most effective strategies and the most critical factors that drive software project turnover. However, more studies must be conducted in other companies and countries to achieve more generalist findings.

Although our study has provided valuable insights, certain limitations must be acknowledged. The research focused on a specific set of countries by merging an interview and survey sample that may not represent the global diversity of the software industry. Further research should be conducted aiming to include a broader and more diverse sample to validate these findings across different cultural and organizational contexts. Additionally, the reliance on self-reported data may introduce bias, as participants' perceptions might be influenced by their current job satisfaction levels. Future studies could benefit from triangulating these findings with quantitative data and longitudinal studies to capture changes over time, which is difficult to execute. New studies could investigate the influence of personal factors, such as personality traits and lifestyle preferences, which could offer a deeper understanding of what motivates Software Engineers to stay or leave their positions. Moreover, examining the impact of remote and hybrid work arrangements on turnover rates could provide relevant insights in the context of the evolving work environment post-pandemic. Finally, research into the role of organizational culture and its interplay with retention strategies can also offer valuable contributions to this field.